\documentclass{aa}
\usepackage{graphics,epsfig}
\usepackage{txfonts}

\newcommand{\grs}    {GRS 1915+105}
\def\simless{\mathbin{\lower 3pt\hbox
     {$\rlap{\raise 5pt\hbox{$\char'074$}}\matHRhar"7218$}}}   
\def\simmore{\mathbin{\lower 3pt\hbox
     {$\rlap{\raise 5pt\hbox{$\char'076$}}\matHRhar"7218$}}}   

\begin{document}

\title{Does \grs\ exhibit "canonical" black-hole states?}

\subtitle{}

\author
{P. Reig \inst{1}, T. Belloni \inst{2}, M. van der Klis \inst{3}
}

\institute{
G.A.C.E, Instituto de Ciencias de los Materiales, University of Valencia,
P.O. Box 22085, E - 46071 Paterna, Valencia, Spain
\and Osservatorio Astronomico di Brera, Via E. Bianchi 46, I-23807 Merate (LC), 
Italy
\and Astronomical Institute ``Anton Pannekoek'', 
University of Amsterdam and Center for High-Energy Astrophysics, 
Kruislaan 403, NL-1098 SJ Amsterdam, the Netherlands
}

\authorrunning{Reig et al.}
\titlerunning{Source states in GRS 1915+105}

\offprints{pablo.reig@uv.es}

\date{Accepted \\
Received : \\
}

\abstract{
We have analysed $RXTE$ data of the superluminal source \grs\ in order to
investigate if, despite its extreme variability, it also exhibits the
canonical source states that characterise other black-hole candidates. The
phenomenology of \grs\ has been described  in terms of three states (named
A, B and C) based on their hardness ratios and position in the
colour-colour diagram. We have investigated the connection between these
states and the canonical behaviour and found that the  shape of the power
spectral continuum  and the values of the best-fit model parameters to the
noise components in all three states indicate that the source shows
properties similar to the canonical very high state. 
 }

\maketitle

\section{Introduction}

Black-hole candidates are observed to go through different states (van der
Klis 1995, Tanaka \& Shibazaki 1996, Liang 1998). The source state is
defined by certain combinations of the value of the model parameters that
are used to account for the X-ray timing and spectral properties.  Since
similar combinations of the spectral and timing parameters are seen in
most black-hole systems the term "canonical" has been coined to designate
those states that are common to many systems (Miyamoto et al. 1992).  The
four canonical states, in addition to quiescence, that have been
recognised are distinguished by the presence or absence of a soft
component, normally modelled as a  multi-colour  black body component  at 
energies  below 5 keV  (Mitsuda et al. 1984 but see Merloni et al. 2000),
by the luminosity and spectral slope of the emission at higher energies
and by the different shapes and characteristic frequencies of the QPO and
noise components. These are the low/hard (LS; Tananbaum et al. 1972),
intermediate (IS; Belloni et al. 1997c, M\'endez and van der Klis 1997,
Belloni et al. 1996), high/soft (HS; Tananbaum et al. 1972) and very high
states (VHS; Miyamoto et al. 1991, Ebisawa et al. 1994).    Recent work
(Rutledge et al. 1999, Homan et al. 2001) has shown that the intermediate
state (IS) and the very high state (VHS)  most likely represent the same
state. Initially, these states were simply  distinguished because in time
the IS occurred in transitions between the low state (LS) and the high
state (HS). The same spectral and timing properties seen at a count rate
above that of the HS in a given source were called VHS properties.
However, since  the VHS can be found at different flux levels there
is no need for two separate states.

\grs\ displays an extraordinary richness in variability, with regular and
stable behaviour interspersed with flaring activity. Belloni et al. (2000,
hereafter B00) analysed a large set of observations of \grs\ and
classified them in 12 separate classes, based on their light curves and
colour-colour diagram (one more class was identified by Klein-Wolt et al.
2002). Despite this complex X-ray behaviour the variability of the source
can be understood as transitions between three basic states:  states A and
B exhibit a soft spectrum, state B having a higher count rate and a
slightly harder spectrum than state A. State C is characterised by a low
count rate and hard spectrum  (B00, see also Markwardt et~al. 1999). The
transitions between these states can be very fast (seconds). States A and
C may appear as short ($<$ 100 s) or long (days) intervals. In contrast,
the source never stays in the B state for more than a few hundreds of
seconds. Ten of the twelve classes are characterised by repeating patterns
of transitions between two, or all three, of the states. There are,
however, two classes, class $\phi$ and $\chi$ which do not show state
transitions (B00). They are characterised by the absence of large
amplitude variations or patterned variability. Class $\phi$ is associated
with state A only and class $\chi$ shows characteristics exclusively of
state C.  The purpose of this paper is to investigate whether we can
establish,  observationally, a relation between the three states observed
in \grs\ and the "canonical" states of other black-hole systems.  A timing
analysis of some of these classes and the implications on various corona
models has recently been presented by Ji et al. (2003).

\begin{table*}
\begin{center}
\caption{Results of the fits. Errors are at 90\% confidence level.}  
\label{res}
\begin{tabular}{cllccccccccc}
\hline\hline \noalign{\smallskip}
Obs.	&Date	&Obs. &State/ &rms$^d$ &$\alpha_1$  &$\nu_{brk}$  &$\alpha_2$ &$\nu_{QPO}$ &FWHM &rms$_{QPO}$  &$\chi^2$(dof) \\
num.	&1996	&ID$^a$ &class&	&		&(Hz)		&	&(Hz)		&(Hz)&	& \\
\noalign{\smallskip} \hline \noalign{\smallskip}
1	&May 21	&I-08-00	&B $\mu$	&9.5  &1.53$\pm$0.03	&-		&-		&-		&- 	&-	&1.49(97) \\
2	&May 26	&I-10-00	&BA$^c$ $\beta$	&9.0  &1.12$\pm$0.03	&5.5$\pm$0.4	&2.1$\pm$0.2	&-		&- 	&- 	&1.67(95) \\
3	&May 29	&I-09-00	&A $\phi$	&7.0  &0.85 fix   	&8.0$\pm$1.0   	&1.4$\pm$0.1	&2.11$\pm$0.05	&2.2$\pm$0.2  &4.1$\pm$0.2 &1.90(119)  \\
4	&May 31	&I-11-00	&A $\phi$	&7.6  &1.08$\pm$0.04	&13.0$\pm$1.0  	&2.4$\pm$0.3	&1.92$\pm$0.03	&1.6$\pm$0.2  &3.4$\pm$0.1 &1.53(106)  \\
5	&Jun 5	&I-12-00	&A $\phi$	&8.9  &1.20$\pm$0.03 	&9.1$\pm$0.7	&2.6$\pm$0.2	&1.57$\pm$0.06	&1.4$\pm$0.2  &3.6$\pm$0.1 &1.36(108)  \\
6	&Jun 7	&I-13-00	&A $\phi$	&9.7  &1.25$\pm$0.08	&8.8$\pm$1.0	&2.7$\pm$0.4	&1.48$\pm$0.06	&1.5$\pm$0.2  &4.3$\pm$0.3 &1.33(108)  \\
7	&Jun 12	&I-14-00	&B $\delta$	&13.1 &1.35$\pm$0.03	&2.2$\pm$0.2	&2.0$\pm$0.1	&-		&-		&- 	&1.72(95)   \\
8	&Jun 22 &I-17-01/02	&A $\phi$	&10.7 &1.34$\pm$0.05	&9.6$\pm$0.6	&-  		&1.20$\pm$0.07	&2.0$\pm$0.2  &5.4$\pm$0.3 &1.48(110)  \\
9	&Jun 25 &I-18-00	&A $\phi$	&10.3 &1.30$\pm$0.07	&8.0$\pm$1.0	&-		&1.40$\pm$0.15	&1.6$\pm$0.5  &4.3$\pm$0.6 &0.84(108)  \\
10	&Jun 29 &I-19-00/01/02	&A $\phi$	&5.3  &1.13$\pm$0.02	&-		&-		&1.98$\pm$0.07	&0.5 fixed	&$<1.2$	   &1.78(114) \\
11	&Jul 3	&I-20-00/01	&A $\phi$	&7.5  &1.07$\pm$0.02	&9.8$\pm$1.2	&1.8$\pm$0.2	&1.66$\pm$0.06	&0.4 fixed	&$<3.9$	   &1.38(145)  \\
12	&Jul 11 &I-22-00	&C $\chi$	&13.3 &0.06$\pm$0.03	&1.72$\pm$0.07  &2.7$\pm$0.2	&3.48$\pm$0.01	&0.40$\pm$0.03	&10.1$\pm$0.2 &1.75(124)  \\
13	&Jul 14 &I-23-00	&C $\chi$	&13.8 &0.16$\pm$0.02	&1.78$\pm$0.09  &2.6$\pm$0.04	&3.56$\pm$0.01	&0.44$\pm$0.03	&10.6$\pm$0.2  &1.98(124)  \\
14	&Jul 16 &I-24-00	&C $\chi$	&15.9 &0.29$\pm$0.02	&2.61$\pm$0.12  &2.72$\pm$0.03	&2.43$\pm$0.01	&0.33$\pm$0.03	&10.9$\pm$0.2  &4.62(124) \\
15	&Jul 19 &I-25-00	&C $\chi$	&17.3 &0.20$\pm$0.05	&2.76$\pm$0.08  &3.12$\pm$0.05	&1.12$\pm$0.01	&0.17$\pm$0.01	&12.2$\pm$0.2  &2.50(124) \\
16	&Jul 26 &I-27-00	&C $\chi$	&16.9 &0.15$\pm$0.02	&1.38$\pm$0.02  &2.53$\pm$0.01	&0.63$\pm$0.01	&0.07$\pm$0.01	&9.5$\pm$0.3   &1.67(124)  \\
17	&Aug 3  &I-28-00	&C $\chi$	&17.4 &0.25$\pm$0.02	&2.12$\pm$0.04  &2.73$\pm$0.01	&0.96$\pm$0.01	&0.12$\pm$0.01	&11.1$\pm$0.2  &2.34(124)  \\
18	&Aug 10 &I-29-00	&C $\chi$	&16.8 &0.17$\pm$0.05	&3.65$\pm$0.05	&3.11$\pm$0.05	&1.85$\pm$0.01	&0.30$\pm$0.03	&12.3$\pm$0.2	&3.33(124)  \\
19	&Aug 18 &I-30-00	&C $\chi$	&13.9 &0.38$\pm$0.04	&2.12$\pm$0.10  &2.11$\pm$0.05	&5.20$\pm$0.02  &1.10$\pm$0.10	&6.6$\pm$0.7	&1.71(122)  \\
20	&Aug 25 &I-31-00	&C $\chi$	&13.7 &0.23$\pm$0.02    &1.59$\pm$0.05	&2.07$\pm$0.06	&3.97$\pm$0.02	&1.09$\pm$0.05	&10.2$\pm$0.2	&2.58(124) \\
21	&Aug 31 &I-32-00	&C $\chi$	&15.9 &0.27$\pm$0.02	&1.55$\pm$0.05	&1.9$\pm$0.1	&5.99$\pm$0.04	&2.15$\pm$0.14	&5.0$\pm$0.1	&5.05(144) \\
22	&Sep 7 &I-33-00		&C $\chi$	&16.5 &0.33$\pm$0.04	&0.91$\pm$0.04	&2.2$\pm$0.2	&5.36$\pm$0.03	&1.14$\pm$0.07	&5.3$\pm$0.1	&3.24(139)  \\
23	&Oct 7 &I-38-00		&B $\lambda$	&6.9  &1.58$\pm$0.06	&-		&-		&-		&-		&-		&1.63(133)  \\
24	&Jun 18$^b$&K-33-00	&B $\kappa$	&13.3  &1.50$\pm$0.02	&-		&-		&-		&-		&-		&1.90(96) \\
25	&Sep 5$^b$&K-45-02	&A $\theta$	&10.9 &1.58$\pm$0.06	&-		&-		&-		&-		&-		&2.32(135)  \\
\noalign{\smallskip} \hline
\multicolumn{8}{l}{$a$: I stands for 10408-01 and K for 20-402-01} \\
\multicolumn{8}{l}{$b$: Year 1997} \\
\multicolumn{8}{l}{$c$: It contains a smooth transition between A and B} \\
\multicolumn{9}{l}{$d$: Obtained from the best-fit model in the frequency range 0.01--20 Hz}  \\
\end{tabular}
\end{center}
\end{table*}
        \begin{figure}
    \begin{center}
    \leavevmode
\epsfig{file=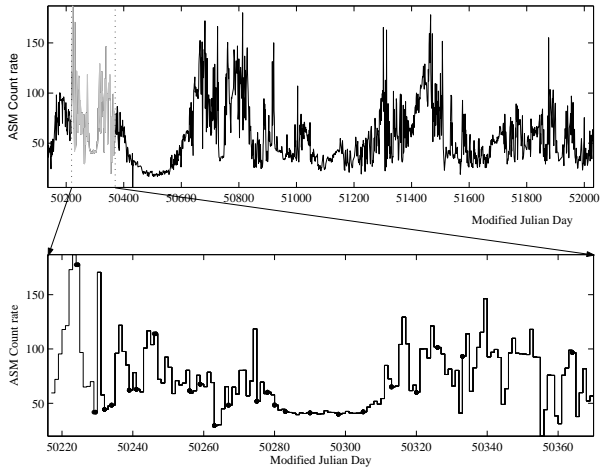, width=8.0cm, clip=}
 \end{center}              
        \caption{ASM light curve of \grs. The period covering our
observations is shown in the lower panel. The dots represent the initial
time of the individual observations}
        \label{asm}
        \end{figure}

\section{Observations}

 We have chosen $RXTE$ PCA observations that represent the three basic
states of B00. These three states occur for various durations in the large
number of variability classes. The fourth column of Table~\ref{res}
indicates from which class the basic states were extracted. Obs. 2
contains a smooth transition between state A and B, virtually impossible to
separate. If one arbitrary cuts the AB interval into two halves and
assumes that one half represents state A and the other half state B then
the results of the spectral fitting for both states agree within
the errors. All three states show variations in count rate by a factor of
two over the time analysed in this work. However, on average state B is
brighter than state C and it, in turn, is brighter than state A. 

Figure \ref{asm} shows the one-day average {\em All Sky Monitor} (ASM)
light curve of \grs. The period covered by the observations is marked by
two dotted lines in the upper panel. An enlargement of that interval is
shown in the lower panel. The dots represent the initial time of the
individual observations.

\begin{figure} 
\begin{center} 
\leavevmode 
\epsfig{file=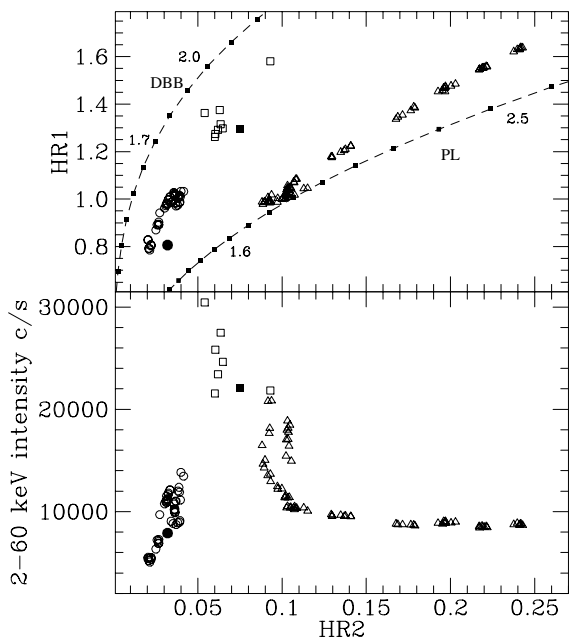, width=8.0cm, bbllx=30pt, bblly=180pt, 
bburx=520pt, bbury=715pt, clip=}
\end{center}               
\caption{Colour-colour and hardness-intensity diagrams. Different states
are represented by different symbols: long state A (open circles),
short state A (filled circle), state B (squares) and state C (triangles).
The filled square represents Obs. 2.
The dashed lines in the CCD  show the positions corresponding to model
spectra that describe the energy spectrum (disc-blackbody and
power law). }
\label{ccd} 
\end{figure}

A soft (HR1) and a hard (HR2) X-ray colour were defined as the ratio of
the intensity 5-13 keV over 2-5 keV and 13-60 keV over 2-5 keV,
respectively.  The colour-colour diagram (CD), shown in the upper panel of
Fig.~\ref{ccd}, clearly separates the observations in three branches which
correspond to each one of the states.  Different symbols represent the
different states: circles for A, squares for B and triangles for C. The
short state A (Obs. 25) is marked by an filled circle  and Obs. 2, which
contains a transition between states A and B by a filled square.  The
dashed lines show the positions corresponding to model spectra that
generally describe the energy spectrum (disk-blackbody and power law,  see
B00 for a description of the CD).  The points along the disc-blackbody
(DBB) line represent steps of 0.1 keV in $kT$. The points along the power
law (PL) line correspond to changes in photon index by steps of 0.1. The
absorption was fixed to $N_H=6.0 \times 10^{22}$ cm$^{-2}$. The pattern
traced by \grs\ in the CD is reminiscent of the comb-like structure
reported by Homan et al. (2001, Fig. 3) for XTE  J1550--564. The main
difference is that the points in the hard branch lie above the power-law
line and not below as in XTE  J1550--564. This could be due to a higher
disk blackbody temperature.

\begin{figure} 
\begin{center} 
\leavevmode 
\epsfig{file=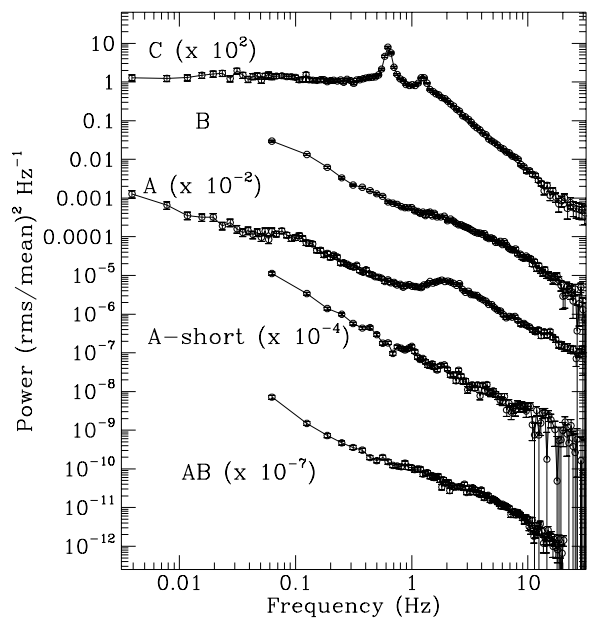,
width=8.0cm, bbllx=45pt, bblly=200pt, bburx=500pt, bbury=685pt, clip=}
\end{center}               
\caption{Representative examples of the three basic states: A (obs. 3), B
(obs. 23), C (obs. 16). Also shown are the short A state (Obs. 25) and the
mixed AB state (Obs. 2)} 
\label{pds} 
\end{figure}

\section{Power spectra}

Power spectra were obtained by dividing the light curves into 256-s
segments and calculating the Fourier spectrum of each segment. The time
resolution was 1/512 seconds, giving a Nyquist frequency of 256 Hz. Given
the shorter duration of the B-state and short A-state observations  16-s
segments were used in those cases. Only photons with energies in the
range 2--13 keV (PCA channels 0--34, PCA epoch 3) were accumulated.
Finally, the power spectra of each segment were averaged together and
logarithmically rebinned to produce one power spectrum for each
observation.

The power spectrum of the observations in the soft branch (state A) is
characterised by a strong red-noise component (Fig.~\ref{pds}). However, a
single power-law function does not fit the data well. Instead, a broken
power law plus one Lorentzian component with frequency in the range 1--3
Hz are required. The peaked noise is normally accompanied by one harmonic.
In Obs. 3-4, an extra broad component at around  0.07 Hz is needed in
order to obtain an acceptable fit. The central frequency of the main
peaked noise component decreased with time from $\sim$ 2 Hz to $\sim$ 1
Hz. The Q-value (the central frequency divided by the  FWHM) was $< 1$,
that is, lower than the generally accepted value of 2 to be considered as
a QPO. The break frequency is always above the peak frequency and varies
in the range 8-13 Hz.  In this state as the 2--13 keV count rate
increased, the fractional root-mean-squared (rms) amplitude in the
frequency range 0.01--20 Hz increased as well.  The power spectrum of
the short A state can be fit by a single power law, i.e., with no
break.  The power spectrum of the B state is qualitatively similar to that
of state A (Fig.~\ref{pds}),  although no broad Lorentzian component is
required.

In the hard branch (state C), the power spectra contain a strong
band-limited noise component together with one QPO peak (occasionally with
some harmonics) in the frequency range 0.01--10 Hz (Fig.~\ref{pds}). The
power spectrum is flat below the QPO frequency and steepens to a power
law above the QPO frequency. As the 2--13 keV count rate increases, the
frequency of the QPO increases and its amplitude (rms) decreases. As the
source becomes harder the rms amplitude increases whereas the QPO frequency
decreases. As in state A a broken power law describing the continuum
and one or two Lorentzians modelling the QPO feature and its harmonics were
used to fit the power spectra. The value of the break frequency is lower
in state C (1--4 Hz) than in state A (8--13 Hz).
Table \ref{res} shows the results of the power spectral fits.  

\begin{table*}
\begin{center}
\caption{Comparison of the spectral and timing parameters of state A, B and C 
of B00 to the canonical low state (LS), high state (HS) and very high state (VHS)}  
\label{comp}
\begin{tabular}{lcccccc}
\hline \hline \noalign{\smallskip}
Parameter	&LS   		&HS		&VHS	&State A  &State B&State C  \\
\noalign{\smallskip} \hline \noalign{\smallskip}
$kT^a$ (keV)	&$<$ 1		&$\sim 1$	&1-2		&1.8	&2.2	&3--4\\
$\Gamma^b$	&1.5--2 	&2--3		&$\sim 2.5$	&$\sim$3.5&$\sim$3.1	&1.5--2.5\\
Noise$^c$	&FT		&PL		&FT \& PL	&PL	&PL	&FT\\
$\nu_{\rm break}$&$\le$ 1 Hz	&-		&$\ge$ 1 Hz&1--3 Hz&-	&$\sim$10 Hz \\
rms		&30--50\%	&$<$ 3\%	&1--15\%	&5--10\%&5--10\%&10--20\%\\
QPO		&no		&no		&1--10 Hz	&no	&no	&1--10 Hz \\
radio		&jet		&no		&bright		&weak	&weak	&jet \\
\noalign{\smallskip} \hline
\multicolumn{7}{l}{$a$: Temperature of the disc blackbody component} \\
\multicolumn{7}{l}{$b$: Power-law index of the hard spectral component} \\
\multicolumn{7}{l}{$c$: Type of noise in the PDS: FT stands for flat-top and PL for power law}  \\
\end{tabular}
\end{center}
\end{table*}

\section{Connection with canonical states}

From the point of view of the timing analysis the power spectra of states
A and B are almost indistinguishable and very similar
to the power-law like power spectrum of other black-hole candidates that 
have gone through the VHS: GX 339-4 (Miyamoto et al. 1991), GS 1124--68
(Miyamoto et al. 1994; Takizawa et al. 1997),  GRO J1655--40 (M\'endez et
al. 1998).  The red- and peaked-noise components and an amplitude of
variability in the range 7--10\% are all properties of the VHS, when the
energy spectrum is soft. As the source becomes harder, i.e, as the source
enters the C state, QPOs in the frequency range 1--10 Hz and flat-top
noise show up in the power spectra. The shape of the power spectra now
resembles that of Cyg X--1 (Belloni et al. 1996), GRO J1655--40 (M\'endez
et al. 1998),  XTE J1748--288 (Revnivtsev et al. 2000) in a IS/VHS. 

The \grs\ observations of June 29 and July 3 present the lowest HR1 ratio
($\sim 0.8$), i.e. they show the softest energy spectrum,  among all
the observations in this paper. In these two observations the source
shows characteristics that approach that of the HS: GRO J1655--40
(M\'endez et al. 1998),  GS 1124--68 (Miyamoto et al. 1993) or GX 339--4
(Belloni et al. 1999), namely, a weak noise component, the absence of QPO,
soft spectrum and red noise in the form of a power law. However, the fact
that the source never reaches the disc blackbody curve (Fig.~\ref{ccd}) and
that the rms amplitude is still rather high (5-7\%) preclude the
association of these observations with the canonical HS.

Although the properties of the radio emission have not been traditionally
used to define the source states it is worth pointing out that  in state
C, \grs\ is a source of strong radio emission (Naik \& Rao 2000; Muno et
al. 2001, Klein-Wolt et al. 2002). In this state the optically thick radio
emission has been shown to originate in a compact jet (Dhawan et al.
2000). Since radio jets in black-hole candidates have been detected in the
LS only (Fender 2001), one may wonder whether state C of \grs\ resembles
the canonical LS. The answer is that neither the  fractional rms, QPO
frequency or break frequency of the flat-top noise component that
characterise state C in \grs\ lie in the range of those found in the
canonical LS: the frequencies are too high and the rms amplitude too low.

\section{Discussion}

\grs\ is unusual among black-hole systems because of its extreme
variability (B00 and references therein) and its long periods
of X-ray activity. \grs\ was discovered as an X-ray transient
(Castro-Tirado et al. 1994) but it has been detected persistently since
then. Permanent X-ray emission is a characteristic of black-hole binaries
with massive companions (Cyg X--1, LMC X--1 and LMC X--3). In contrast,
all confirmed low-mass black-hole binaries are transients. \grs\ contains
a K-MIII star (Greiner et al. 2001).

A proposed model for the interpretation of the X-ray variability involves
the disappearing and subsequent refilling of the inner region of an
accretion disk due to the onset of a thermal/viscous instability (Belloni
et al. 1997a; 1997b). This effect, not observed in other sources on
similarly short time scales, led B00
to define three states, A, B and C based on the X-ray hardness of the
source and its position in the colour-colour diagram. State C was
interpreted as corresponding to the unobservability of the inner portion
of the accretion disc, whereas in  states A and B the disc would be seen
up to the last stable orbit.  The association of the flat-top power
spectrum of \grs\ (state C)  with the canonical VHS has already been
suggested (Morgan et al. 1997; Paul et al. 1997), although the notion of
state C came only later. States A and B have not been explored as such and
their timing association with black hole states is new.

In Table~\ref{comp} a comparison of the timing and spectral parameters of
each state in \grs\ with respect to average values found in other
black-hole systems is presented. The properties of states A and B agree
well with the canonical VHS even though they do not exhibit
proper low-frequency QPOs (in the sense of having a Q-value greater than
2). State C, which does show QPOs, shows a prominent thermal component and
strong radio emission. Also, the power spectrum is power-law like for
states A and B and flat-top like for state C. No correlation is found
between the shape of the noise and the count rate. These relatively rapid
changes (hours to days) of the continuum of the power spectrum of \grs\
have been seen in other systems. During a VHS the power spectrum of GX
339--4 (Miyamoto et al. 1991) and of GS 1124--683 (Miyamoto et al. 1993,
Takizawa et al. 1997) manifested two different shapes according to the
fraction of X-ray photons in the power-law energy spectral component. A
flat-top shape was obtained when there was a substantial amount of counts
contributing to the power-law in the energy spectrum. When this component
was weak the power spectra turned into a power-law shape, that is, the
same behaviour as in \grs.

If one looks into the current models that have been put forward to explain
the phenomenology of the spectral states in black-hole candidates, it is
possible to understand why \grs\ is not so different from other
"well-behaved" systems. Although no current model can explain the whole
phenomenology of this type of systems, it is suggested that in the LS the
accretion disc stops at a certain distance from the black hole, whereas in
the HS the disc extends all the way to the last stable orbit  (Chen \&
Taam 1996; Esin et al. 1997; Zhang et al. 1997). States A and B, being
associated with the observability of the entire disc, can be identified
with high-flux states, i.e., the VHS. Likewise, given that state C
corresponds to the lack of the inner part of the accretion disc, is
perhaps not surprising that it presents some similarities with the
canonical LS. The fact that the canonical HS is not seen in either of
the three states implies that the power-law tail, whose absence
characterises the HS energy spectra of black-hole systems at higher
energies, must always play a strong role in \grs. In short, as in
canonical sources the variations in the value of the innermost radius of
the accretion disc determine the transition between the states. However,
in \grs\ these variations move all in the range of the VHS. 

\section{Conclusion}

The analysis of the aperiodic variability and spectral hardness of \grs\
reveals canonical behaviour but also a number of peculiar properties
compared to other black-hole candidates. The source spends most of its
life in a sort of very high state, showing complex variability.
Occasionally, it transits to a {\em hard}, but not necessarily {\em low}
state where the flux remains stable and no patterned variability is seen,
and where the source can stay for tens of days.  The continuum power
spectra can be described by two different shapes according to  the
hardness of the spectrum: flat-top noise, which appears when the spectrum
is hard, and power-law noise when the spectrum is softer. Such behaviour
has been seen in other black-hole systems during the very high state. No
correlation of the shape of the noise with count rate is seen.

\begin{acknowledgements}

We thank the referee B. Paul for his suggestions that improved the
presentation of the paper. PR is a researcher of the programme  {\em
Ram\'on y Cajal} funded by the Spanish Ministery of Science \& Technology
and by the University of Valencia. The authors also acknowledge partial
support via the European Union Training and  Mobility of Researchers
Network Grant ERBFMRX/CT98/0195. TB acknowledges the Cariplo Foundation
for financial support. This research has made use of data obtained through
the High Energy Astrophysics Science Archive Research Center Online
Service, provided by the NASA/Goddard Space Flight Center.

\end{acknowledgements}

\end{document}